\documentclass[aps,prl,twocolumn,showpacs,superscriptaddress,groupedaddress]{revtex4}  
\usepackage{graphicx}
\usepackage{dcolumn}
\usepackage{amsmath}
\usepackage{bm}        
\usepackage{amssymb}   
\usepackage{multirow}
\newcommand{\etal}{{\em et al}}
 
\begin{document}
 
\title{Comment on ``First Observation of Ground State Dineutron Decay: $^{16}$Be''}

\author{F.M.~Marqu\'es}
\author{N.A.~Orr}
\author{N.L.~Achouri}
\author{F.~Delaunay}
\author{J.~Gibelin}
\affiliation{LPC Caen, ENSICAEN, Universit\'e de Caen, CNRS/IN2P3, 14050 Caen cedex, France}
 
\begin{abstract}{\centerline{\sc Published in Physical Review Letters 109, 239201 (2012).}}\end{abstract}

\maketitle

\begin{figure}
\includegraphics[width=0.48\textwidth]{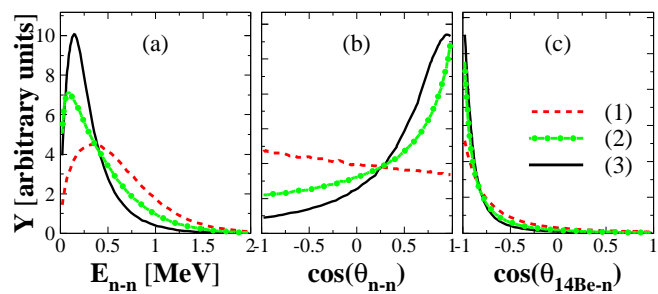}
\caption{\label{Be16-nn-decay} (color online) Calculations of $^{14}$Be+$n$+$n$ decay for
 three-body phase space without (1) and including (2) the $n$-$n$ FSI, and for dineutron decay (3).
 The results have been normalized in each panel to the same integrated yield (Y).} 
\end{figure}
 
 In a recent paper by Spyrou \etal.\ \cite{Spy12}, an investigation of the unbound system $^{16}$Be via single-proton 
 removal from $^{17}$B was reported. In addition to identifying a structure some 1.5~MeV above the $^{14}$Be+$n$+$n$
 threshold, interpreted as the $^{16}$Be ground state, significant enhancements were observed at low $n$-$n$ relative
 energy ($E_{n\mbox{-}n}$) and angle ($\theta_{n\mbox{-}n}$). Through a comparison with simulations based essentially on
 direct three-body and ``dineutron'' decay, Spyrou \etal.\ concluded that only the dineutron mode  
 was consistent with these effects. As such it was claimed that the first case of 
 dineutron decay had been discovered.  Here we point out that such an interpretation is, at best, premature as the
 inclusion of the $n$-$n$ interaction in the description of direct three-body decay can generate strong
 enhancements at low $n$-$n$ relative energy and angle, as observed, without the need to invoke dineutron decay.
 
 An important feature of the interpretation was the treatment of the direct three-body decay mode, whereby
 the well-known $n$-$n$ interaction was neglected. By contrast, the dineutron decay was modeled in terms
 of the two-body decay of $^{16}$Be into $^{14}$Be and a quasi-bound $^2n$ cluster, followed by $n$+$n$ decay.  
 The neglect, however, of the $n$-$n$ interaction in the former case is a significant oversight.
 Indeed, it is well known that the low-energy $^1S_0$ 
 $n$-$n$ interaction invariably leads to a characteristic
 enhancement near zero relative momentum, a feature
 which is exploited in determinations of the $n$-$n$
 scattering length~\cite{Gon06}. More generally, this enhancement is observed in almost any final state in which
 two neutrons are emitted over a relatively short time scale (see, for example, Refs.~\cite{Iek93,FMM00,FMM01,Aum99}).
 It would be surprising, therefore, if such effects were not present in the decay of $^{16}$Be.
  
 To put these considerations on a more quantitative footing, we have undertaken three different calculations (Fig.~1).
 For all three the input is a $^{14}$Be+$n$+$n$ decay-energy 
 distribution following that observed in Ref.~\cite{Spy12}. 
 In case (1) the energy is shared by the three particles following phase space considerations alone.
 In case (3), dineutron decay, it is shared through a sequential process: $^{14}$Be+$^2n$ breakup followed 
 by $^2n$$\rightarrow$$n$+$n$, with
 the $^2n$ decay energy similar to that of Ref.~\cite{Spy12} (Fig.~1a).

 The new case, denoted (2), takes as its starting point three-body phase space which is then modulated by the $n$-$n$ 
 final-state interaction (FSI). The calculations follow the formalism of Ref.~\cite{Led82},
 as described in Ref.~\cite{FMM01}.  For the example shown here an initial $n$-$n$ average separation of $r^{\rm{rms}}_{nn}\!=\!4$~fm
 was assumed. The results are very similar,
 to those of the dineutron model: a clear enhancement is observed at low $n$-$n$ relative angle and energy,
 together with a $^{14}$Be-$n$ distribution peaking sharply at large angles.
 While it is beyond the scope of a Comment to attempt to reproduce exactly the data of Spyrou \etal.,
 if account were taken of the statistics and experimental response function, 
 it would not be possible to discriminate between these two decay processes.
 
 In summary, the inclusion of the $n$-$n$ interaction in the description of direct three-body decay of $^{16}$Be
 generates strong enhancements at low $n$-$n$ relative energy and angle and large $^{14}$Be-$n$ opening angles,
 characteristic of those observed by Spyrou \etal. \cite{Spy12}, without the need to invoke dineutron decay.

\end{document}